\documentclass[12pt]{article}

\pdfoutput=1

\usepackage{graphics}
\usepackage{epsfig}
\usepackage{amsfonts}
\usepackage{amssymb}
\usepackage{amsmath}
\usepackage{dsfont}
\usepackage{pifont}
\usepackage{bbm}
\usepackage{multirow}
\usepackage{latexsym}
\usepackage{verbatim}
\usepackage{mcite}
\usepackage{cite}
\usepackage{units}
\usepackage[all]{xy}
\usepackage{cancel}
\usepackage{slashed}
\usepackage[vcentermath]{youngtab}

\def\a{{\alpha}}

\newcommand{\be}{\begin{equation}}
\newcommand{\ee}{\end{equation}}
\newcommand{\ben}{\begin{displaymath}}
\newcommand{\een}{\end{displaymath}}
\newcommand{\bea}{\begin{eqnarray}}
\newcommand{\eea}{\end{eqnarray}}

\newcommand{\bean}{\begin{eqnarray*}}
\newcommand{\eean}{\end{eqnarray*}}

\DeclareMathAlphabet{\mathpzc}{OT1}{pzc}{m}{it}

\begin{document}
\pagestyle{plain}


\makeatletter \@addtoreset{equation}{section} \makeatother
\renewcommand{\thesection}{\arabic{section}}
\renewcommand{\theequation}{\thesection.\arabic{equation}}
\renewcommand{\thefootnote}{\arabic{footnote}}


\setcounter{page}{1} \setcounter{footnote}{0}


\begin{titlepage}

\begin{flushright}
UUITP-15/15\\
\end{flushright}

\bigskip

\begin{center}

\vskip 0cm

{\LARGE \bf Type IIB on $S^{3}\times S^{3}$ through $Q$ \& $P$ fluxes} \\[6mm]

\vskip 0.5cm

{\bf Ulf Danielsson and\, Giuseppe Dibitetto \,  }\let\thefootnote\relax\footnote{{\tt ulf.danielsson@physics.uu.se, giuseppe.dibitetto@physics.uu.se}}\\

\vskip 25pt

{\em Institutionen f\"or fysik och astronomi, University of Uppsala, \\ Box 803, SE-751 08 Uppsala, Sweden \\[2mm]}

\vskip 0.8cm

\end{center}

\vskip 1cm

\begin{center}

{\bf ABSTRACT}\\[3ex]

\begin{minipage}{13cm}
\small

We study a class of orientifold compactifications of type IIB supergravity with fluxes down to 4D in connection with truncations of half-maximal gauged supergravities yielding isotropic STU-models with 
minimal supersymmetry.
In this context, we make use of a group-theoretical approach in order to derive flux-induced superpotentials for different IIB backgrounds.
We first review the toroidal case yielding GKP-like superpotentials characterised by their \emph{no-scale} behaviour. 
We then turn to $S^{3} \times S^{3}$ and $S^{3} \times \mathbb{T}^{3}$, which, surprisingly, give rise to  effective descriptions of non-geometric $Q$- and $P$-fluxes through globally geometric non-toroidal compactifications.
As a consequence, such constructions break the no-scale symmetry without invoking any non-perturbative effects.

\end{minipage}

\end{center}

\vfill

\end{titlepage}


\tableofcontents

\section{Introduction}
\label{sec:introduction}

In the last two decades a large variety of string compactifications with fluxes have been studied in order to produce lower-dimensional maximally symmetric vacua that might be relevant for cosmological
(de Sitter) or holographic (anti-de Sitter) purposes.

Focusing our attention on type II theories in particular, gauge fluxes (both of NS-NS and R-R type) on a six-torus $\mathbb{T}^{6}$ were introduced as a first ingredient in a compactification scheme generating a
potential for the would-be moduli fields at a perturbative level. 

While this can result in the achievement of full moduli stabilisation in an AdS$_4$ vacuum in massive type IIA with O$6$-planes \cite{DeWolfe:2005uu}, 
in type IIB with O$3$-planes it just produces a class of so-called \emph{no-scale} models \cite{Giddings:2001yu} only allowing for Minkowski solutions where the K\"ahler moduli remain flat. 

Parallelly in ref.~\cite{Derendinger:2004jn}, the idea of including a twisting on the $\mathbb{T}^{6}$ by promoting it to a group manifold with constant spin connection (\emph{a.k.a.} metric flux) was
developed in the context of type IIA orientifold reductions on $\mathbb{T}^{6}/\left(\mathbb{Z}_{2}\times\mathbb{Z}_{2}\right)$ and the connection with $\mathcal{N}=1$ superpotentials in STU-models was 
worked out in detail. 

Conversely, in type IIB, since the option of including metric flux in order to break the aforementioned no-scale symmetry is not available due to its negative parity w.r.t. the orientifold involution,
the possibility of using non-perturbative effects to introduce a dependence on the K\"ahler moduli was initially explored in ref.~\cite{Kachru:2003aw}. However, a possible generic drawback of such 
constructions based on non-perturbative effects, is our lack of information concerning their precise form or their reliability within a supergravity regime. 

A somewhat complementary approach that has been pursued during the last decade is that of introducing \emph{generalised fluxes} \cite{Shelton:2005cf} within the 4D effective description. 
The existence of such objects was originally conjectured on the basis of string duality arguments, though in general no 10D lift is known for these flux deformations.

A strikingly simple and enlightening case for investigating these dual fluxes is that of the $\mathbb{Z}_{2}\times\mathbb{Z}_{2}$ toroidal orbifold of type II compactifications. This is due to the
fact that $\mathbb{T}^{6}/\left(\mathbb{Z}_{2}\times\mathbb{Z}_{2}\right)$ happens to coincide with its own mirror manifold. This implies that different bits of information, which can be accessed in different
duality frames, all find their place in a universal duality-covariant flux-induced superpotential. 

After choosing a specific duality frame, \emph{e.g.} type IIB with O$3$-planes, the majority of the superpotential couplings will represent non-geometric fluxes, \emph{i.e.} with no known higher-dimensional
origin. The aim of the present work is to give evidence for a type IIB lift of some superpotentials generated by non-geometric fluxes of $Q$ \& $P$ type. To do this, we will follow the same 
philosophy as propposed in ref.~\cite{Danielsson:2015tsa}.

The paper is organised as follows. In section~\ref{sec:TypeIIB} we first review orientifold reductions of type IIB supergravity preserving sixteen supercharges in 4D. 
Secondly, we connect these to half-maximal gauged supergravities by showing how turning on fluxes from a \emph{top-down} perspective corresponds to gauging part of the global symmetry of the underlying
4D theory within a \emph{bottom-up} approach.
We will make of use of the aforementioned gauged supergravity theories in their embedding tensor incarnation \cite{Schon:2006kz}. This formulation of gauged supergravity manifestly promotes flux deformations to duality-covariant objects, 
thus containing information concerning dual fluxes. Subsequently we use this formalism as a tool to study the explicit examples of $\mathbb{T}^{6}$, $S^{3}\times S^{3}$ and 
$S^{3}\times\mathbb{T}^{3}$ and derive the embedding tensor/generalised fluxes dictionary. Finally, in section~\ref{sec:Discussion} we speculate on some aspects of our analysis and mention some possible 
future developments.

\section{Type IIB on various compact backgrounds}
\label{sec:TypeIIB}

The low-energy type IIB (pseudo-)action in the string frame reads
\bea
S^{(\textrm{IIB})} & = & \dfrac{1}{(2\pi)^{7}\,(\a^{\prime})^{4}} \, \int d^{10}x \, \sqrt{-g_{10}} \, \left(e^{-2\phi}\mathcal{R}^{(10)} \, + \, 4 e^{-2\phi} (\partial\phi)^{2}
\, - \, \dfrac{1}{2 \cdot 3!}e^{-2\phi} |H_{(3)}|^{2}\right. \, + \notag \\
& & \left. - \,\dfrac{1}{2} |F_{(1)}|^{2}\, - \, \dfrac{1}{2 \cdot 3!} |F_{(3)}|^{2}\, - \, \dfrac{1}{2 \cdot 5!} |F_{(5)}|^{2}\right) \, + \, \textrm{C-S} \ , \label{IIB_action}
\eea
where $F_{(5)}$ satisfies the following self-duality condition $F_{(5)} \, \overset{!}{=} \, \star_{10}F_{(5)}$.

We choose the following reduction \emph{Ansatz}
\be
\label{IIB_red_ansatz}
ds^{2}_{10} \, = \, \tau^{-2} \, ds^{2}_{4} \, + \, \rho \, g_{mn} \, dy^{m} \,\otimes\, dy^{n} \ ,
\ee
where $\tau$ and $\rho$ are suitable combinations of the internal volume $\textrm{vol}_{6}$ and the ten-dimensional dilaton $\phi$ which are usually referred to as the universal moduli
\cite{Hertzberg:2007wc}. The internal geometry is parametrised by the element $g_{mn}$ of the $\textrm{SL}(6)/\textrm{SO}(6)$ coset.
According to \eqref{IIB_red_ansatz}, the ten-dimensional Ricci scalar $\mathcal{R}^{(10)}$ reduces to
\be
\label{Ricci_red}
\begin{array}{cccc}
\mathcal{R}^{(10)} & \longrightarrow & \tau^{2} \, \mathcal{R}^{(4)} \, + \, \rho^{-1} \, \mathcal{R}^{(6)} & .
\end{array}
\ee
Imposing
\be
\label{Eistein_4D}
e^{2\phi} \, = \, \tau^{-2}\rho^{3}
\ee
guarantees a four-dimensional Lagrangian in the Einstein frame. By performing the dimensional reduction of the various kinetic terms in the action \eqref{IIB_action}, one can derive the $(\rho,\tau)$ scaling of the corresponding fluxes in a very
straightforward way. 

As an example, from a reduction of the corresponding term in \eqref{IIB_red_ansatz}, one finds that the $(\rho,\tau)$ weights of $F_{mnp}$ are
\be
\begin{array}{lclclc}
\sqrt{-g_{10}} \, |F_{(3)}|^{2} & \longrightarrow & \tau^{-4}\rho^{3} \, |F_{mnp}|^{2} \, \rho^{-3} & = &
\tau^{-4} \, |F_{mnp}|^{2} & ,
\end{array}
\label{F_scaling}
\ee
where $|F_{mnp}|^{2} \, \equiv \, F_{mnp}F_{qrs}g^{mq}g^{nr}g^{ps}$. 

\subsection{Orientifold reductions of type IIB supergravity}
\label{sec:Orientifold}

In the presence of O$3$-planes, the 10D field content undergoes a truncation that selects the even sector w.r.t. to a combination of worldsheet parity $\Omega_{p}$, fermionic number $(-1)^{F_{L}}$ and 
orientifold involution. From a world-sheet perspective, \emph{i.e.} under the combined $\,(-1)^{F_{L}}\Omega_{p}\,$ action, the type IIB fields $g \,,\, \phi \,,\, C_{(0)}$ and $C_{(4)}$ are parity-even 
whereas $B_{(2)}$ and $C_{(2)}$ are parity-odd.

In our compactifications we will consider O-planes placed as follows
\be
\begin{array}{lcccc}
\textrm{O$3$-planes} \, : & & & \underbrace{\times \, \vert \, \times \, \times \, \times}_{D = 4} \, \, \underbrace{- \, - \, -\, - \, - \, -}_{m} & ,
\end{array}
\notag
\ee
and subsequently define the associated orientifold involution by
\be
\label{sigma_O3}
\sigma_{\textrm{O}3} \,\,: \,\, (\, y^{1} \, , \, y^{2} \, , \, y^{3} \, , \, y^{4} \, , \, y^{5} \, , \,y^{6} \, ) 
\,\, \rightarrow \,\, 
(\, -y^{1} \, , \, -y^{2} \, , \, -y^{3} \, , \, -y^{4} \, , \, -y^{5} \, , \, -y^{6} \, ) \ .
\ee
The above conventions automatically assign a $\mathbb{Z}_{2}$ parity to the six physical coordinates on the internal manifold $X_{6}$ that is induced by the O$3$-involution in \eqref{sigma_O3}.  
\be
\begin{array}{lclclc}
x^{M} & \longrightarrow & \underbrace{x^{\mu}}_{\textrm{4D}} & \oplus & \underbrace{y^{a} \,\oplus\, y^{i}}_{(-)} & ,
\end{array}\label{parity}
\ee
where $y^{m}\,\equiv\,\left(y^{a}, \, y^{i}\right)$ realise the compact geometry of $X_{6}$. Retaining only even fields and fluxes w.r.t. the action of the above $\mathbb{Z}_{2}$ will automatically restrict
our supergravity theory obtained upon such a type IIB reduction within the framework of $\mathcal{N}=1$ STU-models.   

In order to identify all the three scalar excitations within the aforementioned STU-models, we need to open up an extra \emph{semi-universal} deformation of the metric \eqref{IIB_red_ansatz}. This
yields the following new 10D Ansatz
\be
\label{IIB_nonuniv_ansatz}
ds^{2}_{10} \, = \, \tau^{-2} \, ds^{2}_{4} \, + \, \rho \, \left(\sigma^{-1} \, g_{ab} \, \eta^{a} \otimes \eta^{b} \, + \, \sigma \, g_{ij} \, \eta^{i} \otimes \eta^{j} \right) \ ,
\ee
where $\left\{\eta^{m}\right\}\,\equiv\,\left\{\eta^{a}, \, \eta^{i}\right\}$ represent a basis of one-forms carrying the information about the dependence of the metric on the internal coordinates.  
 The extra $\mathbb{R}^{+}$ scalar $\sigma$ parametrises the relative size between the $a$ and $i$ coordinates, which would acquire opposite involution-parity when adopting the type IIA picture 
\cite{Danielsson:2012et}. 
Moreover, $g_{ab}$ and $g_{ij}$ contain in general $\textrm{SL}(3)_{a} \, \times \, \textrm{SL}(3)_{i}$ scalar excitations. However, we will keep such degrees of freedom frozen here by imposing the 
extra requirement of $\textrm{SO}(3)$-invariance, \emph{i.e.} $g_{ab} \, = \, \delta_{ab}$ and $g_{ij} \, = \, \delta_{ij}$. This will constructively yield an \emph{isotropic} STU-model in 4D.

The relationship between the STU scalars and the geometric moduli appearing in \eqref{IIB_nonuniv_ansatz} reads
\be
\label{rhotausigma/STU}
\begin{array}{lclclc}
\rho \ = \ \textrm{Im}(S)^{-1/2} \, \textrm{Im}(T)^{1/2} & , &  \tau \ = \ \textrm{Im}(S)^{1/4} \, \textrm{Im}(T)^{3/4} & , & 
\sigma \ = \ \textrm{Im}(U) & .
\end{array}
\ee
The STU-scaling weights, and the $\mathbb{Z}_{2}$-parity of all type IIB fields, were already worked out in ref.~\cite{Dibitetto:2014sfa}. In table~\ref{10DFields} we summarise and recollect the results of the analysis done there.
\begin{table}[t!]
\begin{center}
\scalebox{0.80}[0.85]{
\begin{tabular}{|c||c|c|c|c|}
\hline
B/F & $\sigma_{\textrm{O}3}$ & $(-1)^{F_{L}}\,\Omega_{p}$ & IIB field & $\textrm{SL}(3)_{a} \, \times \, \textrm{SL}(3)_{i} \times \mathbb{R}^{+}_{S} \times \mathbb{R}^{+}_{T}\times \mathbb{R}^{+}_{U}$ \\[1mm]
\hline\hline
\multirow{13}{*}{B} & $+$ & $+$ & $\phi$ & $(\textbf{1},\textbf{1})_{(0;\,0;\,0)}$ \\[2mm]
\cline{2-5} & $+$ & $+$ & ${e_{a}}^{a}\,=\,{e_{i}}^{i}$ & $(\textbf{1},\textbf{1})_{(0;\,0;\,0)}$ \\[2mm]
\cline{2-5} & $+$ & $+$ & ${e_{a}}^{b}$ & $(\textbf{8},\textbf{1})_{(0;\,0;\,0)}$ \\[2mm]
\cline{2-5} & $+$ & $+$ & ${e_{i}}^{j}$ & $(\textbf{1},\textbf{8})_{(0;\,0;\,0)}$ \\[2mm]
\cline{2-5} & $+$ & $+$ & ${e_{a}}^{i}$ & $(\textbf{3}^{\prime},\textbf{3})_{(0;\,0;\,-1)}$ \\[2mm]
\cline{2-5} & $+$ & $+$ & ${e_{i}}^{a}$ & $(\textbf{3},\textbf{3}^{\prime})_{(0;\,0;\,+1)}$ \\[2mm]
\cline{2-5} & $+$ & $+$ & ${e_{m}}^{m}$ & $(\textbf{1},\textbf{1})_{(0;\,0;\,0)}$ \\[2mm]
\cline{2-5} & $+$ & $+$ & $C_{(0)}$ & $(\textbf{1},\textbf{1})_{(+1;\,0;\,0)}$ \\[2mm]
\cline{2-5} & $+$ & $+$ & $C_{aijk}$ & $(\textbf{3}^{\prime},\textbf{1})_{(0;\,+1;\,+1)}$ \\[2mm]
\cline{2-5} & $+$ & $+$ & $C_{abci}$ & $(\textbf{1},\textbf{3}^{\prime})_{(0;\,+1;\,-1)}$ \\[2mm]
\cline{2-5} & $+$ & $+$ & $C_{abij}$ & $(\textbf{3},\textbf{3})_{(0;\,+1;\,0)}$ \\[2mm]
\hline\hline
\multirow{10}{*}{F} & $+$ & $-$ & $B_{ab}$ & $(\textbf{3},\textbf{1})_{(-\frac{1}{2};+\frac{1}{2};\,+1)}$ \\[2mm]
\cline{2-5} & $+$ & $-$ & $B_{ij}$ & $(\textbf{1},\textbf{3})_{(-\frac{1}{2};+\frac{1}{2};\,-1)}$ \\[2mm]
\cline{2-5} & $+$ & $-$ & $B_{ai}$ & $(\textbf{3}^{\prime},\textbf{3}^{\prime})_{(-\frac{1}{2};+\frac{1}{2};\,0)}$ \\[2mm]
\cline{2-5} & $+$ & $-$ & $B_{abcijk}$ & ($\textbf{1},\textbf{1})_{(-\frac{1}{2};-\frac{3}{2};\,0)}$ \\[2mm]
\cline{2-5} & $+$ & $-$ & $C_{ab}$ & $(\textbf{3},\textbf{1})_{(+\frac{1}{2};+\frac{1}{2};\,+1)}$ \\[2mm]
\cline{2-5} & $+$ & $-$ & $C_{ij}$ & $(\textbf{1},\textbf{3})_{(+\frac{1}{2};+\frac{1}{2};\,-1)}$ \\[2mm]
\cline{2-5} & $+$ & $-$ & $C_{ai}$ & $(\textbf{3}^{\prime},\textbf{3}^{\prime})_{(+\frac{1}{2};+\frac{1}{2};\,0)}$ \\[2mm]
\cline{2-5} & $+$ & $-$ & $C_{mnpqrs}$ & $(\textbf{1},\textbf{1})_{(+\frac{1}{2};-\frac{3}{2};\,0)}$ \\[2mm]
\hline
\end{tabular}
}
\end{center}
\caption{{\it The physical scalars from type IIB compactifications mapped into states in the decomposition of the $\textbf{133}$ of $\textrm{E}_{7(7)}$, \emph{i.e.} the U-duality group in 4D. 
Note that it is the combination $(-1)^{F_{L}}\,\Omega_{p}\,\sigma_{\textrm{O}3}$ of fermionic number, 
worldsheet parity and orientifold involution what determines which states are ``bosonic'' (B) (kept) and ``fermionic'' (F) (projected out). }}
\label{10DFields}
\end{table}

In the second part of this section we will be considering some examples of orientifold reductions of type IIB supergravity with O$3$-planes leading to STU-models within $\mathcal{N}=1$ supergravity in 4D.
For each of them we will propose a group-theoretical derivation of the corresponding flux-induced superpotential which follows the prescription adopted in ref.~\cite{Danielsson:2015tsa} in the context of
M-theory reductions. 

We will start out by revisiting the case of $\mathbb{T}^{6}$ compactifications giving rise to GKP-like backgrounds \cite{Giddings:2001yu} and we will derive the flux-induced superpotential for 
this class of theories through the aforementioned group-theoretical considerations. This will help us construct th working conventions to be used in the analogous derivation that will be carried out 
for different non-toroidal backgrounds. Before we do this, we need to first introduce a particular group-theoretical truncation of half-maximal supergravity in 4D leading to the isotropic STU-models that 
we are interested in. 
e
\subsection{An $\textrm{SO}(3)$ truncation of $\mathcal{N}=4$ supergravity}
\label{sec:truncation}

Half-maximal supergravity in 4D coupled to six vector multiplets arises from $\mathbb{T}^{6}$ reductions of orientifolds of type II theories. It enjoys $\textrm{SL}(2) \times \textrm{SO}(6,6)$ global 
symmetry and all its fields and deformations (\emph{i.e.} gaugings) transform in irrep's of such a global symmetry group \cite{Schon:2006kz}.

Starting out from $\mathcal{N}=8$ supergravity in 4D, and proceeding in a somewhat ``bottom-up'' way, the orientifold involution described in section~\ref{sec:Orientifold} may be viewed as the following 
$\mathbb{Z}_{2}$ truncation (see \eqref{E7->SL2SO66})
\be
\begin{array}{ccl}
\textrm{E}_{7(7)} & \supset & \textrm{SL}(2)_{S} \times \textrm{SO}(6,6) \ , \\[2mm]
\textbf{56} & \overset{\mathbb{Z}_{2}}{\rightarrow} & (\textbf{2},\textbf{12})_{(+)} \oplus (\textbf{1},\textbf{32})_{(-)} \ ,
\end{array}
\nonumber
\ee
which retains its \emph{even} sector, thus breaking half of the original supersymmetry. This procedure yields (gauged) $\mathcal{N}=4$ supergravity in $D=4$ \cite{Dibitetto:2011eu}.

In particular, the vector fields of the half-maximal theory transform in the $(\textbf{2},\textbf{12})$ though only half of them are physically independent due to 4D electromagnetic duality, the scalar 
fields transform in the $(\textbf{3},\textbf{1})\,\oplus\,(\textbf{1},\textbf{66})$ though only $2 \, + \, 36 \, = \,38$ of them are physically propagating due to the presence of a local 
$\textrm{SO}(2)\times\textrm{SO}(6)\times\textrm{SO}(6)$ symmetry. 

A group-theoretical truncation consists in branching all fields and deformations of the theory into irrep's of a suitable subgroup $G_{0} \subset \textrm{SL}(2)_{S} \times \textrm{SO}(6,6)$ and 
retaining only the $G_{0}$-singlets. Such a truncation is guaranteed to be mathematically consistent due the covariance of the eom's of half-maximal supergravity w.r.t. its global symmetry. More precisely
said, $G_{0}$-singlets can only source the eom's of other singlets, thus making it possible to consistently get rid of all the non-singlet modes.

In this context, we need to perform the correct truncation that makes contact with the $\mathcal{N}=1$ isotropic STU-models mentioned in section~\ref{sec:Orientifold} providing an effective description of 
orientifold compactifications of type IIB supergravity down to 4D.  
Such a suitable truncation turns out to be the one retaining the $\textrm{SO}(3)$-invariant sector of half-maximal supergravity, \emph{i.e.}
\be
\textrm{SL}(2)_{S} \times \textrm{SO}(6,6) \,\,\supset \,\,\textrm{SL}(2)_{S} \times \textrm{SO}(2,2) \times \textrm{SO}(3) \,\approx\,  \prod_{\Phi=S,T,U} \textrm{SL}(2)_{\Phi} \,\times\, \textrm{SO}(3) \ .
\ee
This step breaks half-maximal to minimal $\,\mathcal{N}=1\,$ supergravity due to the decomposition $\,\textbf{4} \rightarrow \textbf{1} \oplus \textbf{3}\,$ of the fundamental representation of the 
$\,\textrm{SU}(4)\,$ R-symmetry group in $\mathcal{N}=4$ supergravity under the $\,\textrm{SO}(3)\,$ subgroup
\be
\textrm{SU}(4)_{R} \,\,\supset \,\, \textrm{SU}(3) \,\,\supset \,\, \textrm{SO}(3) \ .
\ee
The resulting theory does not contain any vectors since there are no $\,\textrm{SO}(3)$-singlets in the decomposition $\,\textbf{12} \rightarrow (\textbf{4},\textbf{3})\,$ of the fundamental representation of 
$\,\textrm{SO}(6,6)\,$ under $\,\textrm{SO}(2,2) \times \textrm{SO}(3)$. The physical scalar fields span the coset space
\be
\mathcal{M}_{\textrm{scalar}}= \prod_{\Phi=S,T,U}\left( \frac{\textrm{SL}(2)}{\textrm{SO}(2)} \right)_{\Phi} \ ,
\ee
involving three $\,\textrm{SL}(2)/\textrm{SO}(2)\,$ factors each of which can be parameterised by a complex scalar $\,\Phi=(S , T , U)$. The explicit embedding of the $\mathcal{N}=1$ scalars within the
$38$ scalars of the $\mathcal{N}=4$ theory reads \cite{Dibitetto:2011gm}
\be
M_{\alpha\beta} \ = \ \frac{1}{\textrm{Im}(S)}\,\left(\begin{array}{cc}
|S|^{2} & \textrm{Re}(S) \\
\textrm{Re}(S) & 1 
\end{array}\right)\ \in \ \left( \frac{\textrm{SL}(2)}{\textrm{SO}(2)} \right)_{S} \ ,
\ee
and 
\be
M_{MN} \ = \ \left(\begin{array}{cc}
G^{-1} & -G^{-1}\,B \\
B\,G^{-1} & G\,-\,B\,G^{-1}\,B 
\end{array}\right) \, \otimes \, \mathds{1}_{3} \ \in \  \frac{\textrm{SO}(6,6)}{\textrm{SO}(6) \times \textrm{SO}(6)}  \ ,
\ee
where
\be
\begin{array}{lcclc}
G \ \equiv \ \frac{\textrm{Im}(T)}{\textrm{Im}(U)}\,\left(\begin{array}{cc}
|U|^{2} & -\textrm{Re}(U) \\
-\textrm{Re}(U) & 1 
\end{array}\right) & , & \textrm{and} & B \ \equiv \ \left(\begin{array}{cc}
0 & \textrm{Re}(T) \\
-\textrm{Re}(T) & 0 
\end{array}\right) & .
\end{array}
\ee

The kinetic Lagrangian of this sector can be effectively derived from the following K\"ahler potential 
\be
K \ = \ -\log\left(-i\,(S-\bar{S})\right) \, - \, 3\log\left(-i\,(T-\bar{T})\right) \, - \, 3\log\left(-i\,(U-\bar{U})\right) \ .
\ee

The unimodular deformations (\emph{i.e.} gaugings) of the theory, which are encoded by the so-called \emph{embedding tensor}, transform in the $(\textbf{2},\textbf{220})$ and can be arranged into
an object denoted by $f_{\alpha [MNP]}$ \cite{Schon:2006kz}.

When performing the $\textrm{SO}(3)$ truncation, the embedding tensor reduces to a set of $\,40\,$ invariant components 
\be
\label{ET_splitting}
\begin{array}{lclc}
f_{\alpha [MNP]} & \longrightarrow & \underbrace{\Lambda_{\alpha (ABC)}}_{\left(\textbf{2},\,\tiny{\yng(3)}\,\right) \textrm{ of } \textrm{SL}(2)\times\textrm{SO}(2,2)} \ \otimes \ 
\underbrace{\epsilon_{IJK}}_{\textbf{1} \textrm{ of } \textrm{SO}(3)} & ,
\end{array}
\ee
which can be viewed as the superpotential couplings\footnote{The connection between the $\,\mathcal{N}=1\,$ and
$\mathcal{N}=4$ theory was extensively investigated in ref.~\cite{Aldazabal:2008zza}. However, the explicit agreement between the  scalar potentials up to quadratic constraints was first shown in 
ref.~\cite{Dibitetto:2011gm}.} representing a complete duality-inviariant set of \emph{generalised fluxes} \cite{Shelton:2005cf}. 
This yields the following duality-covariant flux-induced superpotential 
\be
\label{W_fluxes4}
\mathcal{W} \ = \ (P_{F} - P_{H} \, S ) + 3 \, T \, (P_{Q} - P_{P} \, S ) + 3 \, T^2 \, (P_{Q'} - P_{P'} \, S ) + T^3 \, (P_{F'} - P_{H'} \, S ) \ ,
\ee
involving the three complex moduli $S$, $T$ and $U$ surviving the $\textrm{SO}(3)$ truncation introduced ealier in this section. 
\be
\begin{array}{lcll}
\label{Poly_unprim}
P_{F} = a_0 - 3 \, a_1 \, U + 3 \, a_2 \, U^2 - a_3 \, U^3 & \hspace{5mm},\hspace{5mm} & P_{H} = b_0 - 3 \, b_1 \, U + 3 \, b_2 \, U^2 - b_3 \, U^3 & ,  \\[2mm]
P_{Q} = c_0 + C_{1} \, U - C_{2} \, U^2 - c_3 \, U^3 & \hspace{5mm},\hspace{5mm} & P_{P} = d_0 + D_{1} \, U - D_{2} \, U^2 - d_3 \, U^3 & ,
\end{array}
\ee
as well as those induced by their primed counterparts $\,(F',H')\,$ and $\,(Q',P')\,$ fluxes \cite{Aldazabal:2006up},
\be
\begin{array}{lcll}
\label{Poly_prim}
P_{F'} = a_3' + 3 \, a_2' \, U + 3 \, a_1' \, U^2 + a_0' \, U^3 & \hspace{3mm},\hspace{3mm} &P_{H'} = b_3' + 3 \, b_2' \, U + 3 \, b_1' \, U^2 + b_0' \, U^3 & ,  \\[2mm]
P_{Q'} = -c_3' +  C'_{2} \, U + C'_{1} \, U^2 - c_0' \, U^3 & \hspace{3mm},\hspace{3mm} & P_{P'} = -d_3' + D'_{2} \, U + D'_{1} \, U^2 - d_0' \, U^3 & .
\end{array}
\ee
For the sake of simplicity, we have introduced the flux combinations $\,C_{i} \equiv 2 \, c_i - \tilde{c}_{i}\,$, $\,D_{i} \equiv 2 \, d_i - \tilde{d}_{i}\,$, $\,C'_{i} \equiv 2 
\, c'_i - \tilde{c}'_{i}\,$ and $\,D'_{i} \equiv 2 \, d'_i - \tilde{d}'_{i}\,$ entering the superpotential (\ref{W_fluxes4}), and hence also the scalar potential.

For more details concerning the physical interpretation of the above embedding tensor deformations and type IIB orientifold-even generalised fluxes, we refer to appendix~\ref{sec:app2}.

\subsection{Tadpoles and quadratic constraints}
\label{sec:QC}

In the previous section we have spelled out some details concerning the correspondence between embedding tensor deformations $f_{\alpha MNP}$ of the half-maximal 4D theory and orientifold-even 
generalised type IIB fluxes. Such an analysis results in the dictionary in tables~\ref{table:unprimed_fluxes4} and \ref{table:primed_fluxes4}. However, on the gauged supergravity side, the components
of $f_{\alpha MNP}$ only describe a consistent $\mathcal{N}=4$ gauging provided that the following set of \emph{quadratic constraints} (QC) is satisfied
\be
\label{QC4}
\begin{array}{lclclc}
\textrm{QC}_{4} & : & {f_{\alpha [MN}}^{R} \, f_{\beta PQ]R} \ = \ 0 & , & \epsilon^{\alpha\beta} \, {f_{\alpha MN}}^{R} \, f_{\beta PQR} \ = \ 0 & ,
\end{array}
\ee
ensuring the closure of the gauge algebra.

If one furthermore wants to demand the existence of an uplift of the above gaugings to the maximal theory, the following two extra QC are needed \cite{Dibitetto:2011eu}
\be
\label{QC8}
\begin{array}{lclclc}
\textrm{QC}_{8} & : &\epsilon^{\alpha\beta} \,f_{\alpha [MNP} \, f_{\beta QRS]} \ = \ 0 & , &  f_{\alpha MNP} \, {f_{\beta}}^{MNP} \ = \ 0 & .
\end{array}
\ee

When retranslating the components of the embedding tensor back into generalised fluxes, the above sets of QC represent nothing but \emph{tadpole} conditions enforcing the absence of SUSY-breaking
extended sources. These would be inconsistent with the amount of supercharges possessed by the original theory.

So, in particular, the QC in \eqref{QC4} are required for consistency of the half-maximal theory and, as such, they set to zero all the flux tadpoles which would need to be cancelled by extended objects
breaking supersymmetry further down to $\mathcal{N}<4$. Conversely, all those other tadpoles which can be sourced by BPS branes preserving the same sixteen supercharges will be left arbitrary by the 
\eqref{QC4}.

On the other hand, the absence of all of the latter tadpoles will be required by the extra QC in \eqref{QC8}, which are needed for the existence of an $\mathcal{N}=8$ lift. 

In summary, whenever studying a candidate embedding tensor configuration to describe an orientifold of type IIB, the QC \eqref{QC4} should be satisfied, whereas the non-zero rhs of \eqref{QC8} will tell
us about the type of BPS local sources that support the string background in question. The general situation can be therefore depicted as follows
\begin{center}
\scalebox{1}[1]{\xymatrix{
*+[F-,]{\begin{array}{c}\boldsymbol{\textbf{Gauged SUGRA}}\\
f_{\alpha MNP} \\ \textrm{QC}_{4} \ \overset{!}{=} \ 0 \\ \textrm{QC}_{8} \ \neq \ 0\end{array} } & \longleftrightarrow &
*+[F-,]{\begin{array}{c}\textbf{Fluxes}\\
\left\{a_{0},\,\dots,\,d_{3}'\right\} \\ \textrm{non-BPS branes} \\ \textrm{BPS branes} \end{array} } & , }}
\end{center}
where, in the above picture, the type IIB fluxes are generically \emph{generalised} (\emph{i.e.} U-dual) \cite{Aldazabal:2010ef} and the corresponding branes are, as a consequence, 
\emph{exotic} \cite{Bergshoeff:2012ex,deBoer:2012ma}.

\subsection{Compactifications on $\mathbb{T}^{6}$}
\label{sec:T6}

In the type IIB toroidal case with O$3$-planes, the requirement of $\textrm{SO}(3)$-invariance truns out to be equivalent to performing an isotropic $\mathbb{Z}_{2} \times \mathbb{Z}_{2}$ orbifold projection.
Hence it is possible to turn on both NS-NS and R-R 3-form gauge fluxes, whereas the orientifold projection together with the $\mathbb{Z}_{2} \times \mathbb{Z}_{2}$ orbifold action eliminate 1- and 5-form 
gauge fluxes as well as the possibility of twisting the $\mathbb{T}^{6}$ by adding metric flux.

Such GKP-like backgrounds, which were originally studied in ref.~\cite{Giddings:2001yu}, are generically supported by the presence of D$3$-branes and O$3$-planes and hence they are described by means of
a gauged $\mathcal{N}=4$ supergravity in 4D. By restricting oneself to the isotropic sector (see section~\ref{sec:truncation}), such theories admit an $\mathcal{N}=1$ description within an STU-model.

In order to identify the emebedding tensor/fluxes dictionary, we need to branch the object $f_{\alpha MNP}$ w.r.t. the following chain of maximal subgroups\footnote{One could 
have made the following alternative choice 
\be
\begin{array}{lclclc}
\textrm{SL}(2)\times\textrm{SO}(6,6) & \supset & \mathbb{R}^{+}_{\Sigma}\times\mathbb{R}^{+}_{1}\times\textrm{SL}(6) & \supset & 
\mathbb{R}^{+}_{\Sigma}\times\mathbb{R}^{+}_{1}\times\mathbb{R}^{+}_{2}\times\textrm{SL}(3)_{a}\times\textrm{SL}(3)_{i} & ,
\end{array}\notag
\ee
which appears to be more natural for the $\mathbb{T}^{6}$ case. However, the decompostion chain used here is the natural one for the cases that will be presented in the next subsections.
Moreover, we note here that the two aforementioned different branching routes yield the same final result up to a relabelling of the three $\mathbb{R}^{+}$ weights.
}
\be
\begin{array}{lclclc}
\textrm{SL}(2)\times\textrm{SO}(6,6) & \supset & \mathbb{R}^{+}_{\Sigma}\times\textrm{SL}(4)_{a}\times\textrm{SL}(4)_{i} & \supset & 
\mathbb{R}^{+}_{\Sigma}\times\mathbb{R}^{+}_{a}\times\mathbb{R}^{+}_{i}\times\textrm{SL}(3)_{a}\times\textrm{SL}(3)_{i} & ,
\end{array}\notag
\ee
where now the two $\textrm{SL}(3)$ factors realise the six physical internal coordinates. Since all internal directions are orientifold-odd, the physical derivative operators are found within the 
decomposition of the $\textbf{32}$ (\emph{i.e.} spinorial) irrep of $\textrm{SO}(6,6)$. This yields (see appendix~\ref{sec:app1})
\be
\begin{array}{cclc}
\textrm{SL}(2)\times\textrm{SO}(6,6)  & \supset  & \mathbb{R}^{+}_{\Sigma}\times\mathbb{R}^{+}_{a}\times\mathbb{R}^{+}_{i}\times\textrm{SL}(3)_{a}\times\textrm{SL}(3)_{i} & , \\[3mm]
(\textbf{1},\textbf{32})  & \rightarrow  & \underbrace{(\textbf{3}',\textbf{1})_{(0;-1;-3)}}_{\partial_{a}}  \oplus \underbrace{(\textbf{1},\textbf{3}')_{(0;-3;-1)}}_{\partial_{i}} 
\oplus \dots & .
\end{array}\label{Derivative_branching}
\ee
Please note that all the examples of flux backgrounds studied in this paper only retain deformations that can be constructed as states obtained by acting with the physical derivatives in \eqref{Derivative_branching}
on some of the internal components of the gauge fields listed in table~\ref{10DFields}, thus yielding by construction locally geometric backgrounds in the toroidal sense.
 
According to \cite{Dibitetto:2014sfa}, the internal derivative operators should correspond to the STU states $(\textbf{3}',\textbf{1})_{(0;+1;-\frac{1}{2})}$ and 
$(\textbf{1},\textbf{3}')_{(0;+1;+\frac{1}{2})}$, respectively. This, together with a suitable normalisation of $\mathbb{R}^{+}_{\Sigma}$, uniquely determines the following mapping between
the STU-weights and the $\mathbb{R}^{+}$-weights labelled by ``$\Sigma$'', ``$a$'' and ``$i$'' associated with the conventions in appendix~\ref{sec:app1}
\be
\left\{
\begin{array}{lclc}
q_{S} & = & \frac{1}{2} \, q_{\Sigma} & , \\
q_{T} & = & -\frac{1}{4} \, \left(q_{a}\,+\,q_{i}\right) & , \\
q_{U} & = & -\frac{1}{4} \, \left(q_{a}\,-\,q_{i}\right) & .
\end{array}
\right.
\label{dictionary_charges}
\ee

Moving to the fluxes, we decompose the $(\textbf{2},\textbf{220})$ into
\be
\begin{array}{cclc}
(\textbf{2},\textbf{220})  & \rightarrow  &  (\textbf{1},\textbf{1})_{(+1;-6;0)}  \oplus (\textbf{1},\textbf{1})_{(-1;-6;0)} \oplus 
(\textbf{3}',\textbf{3})_{(+1;-4;-2)}  \oplus (\textbf{3}',\textbf{3})_{(-1;-4;-2)}  \oplus \\
 & & (\textbf{3},\textbf{3}')_{(+1;-2;-4)}  \oplus (\textbf{3},\textbf{3}')_{(-1;-2;-4)} \oplus 
(\textbf{1},\textbf{1})_{(+1;0;-6)}  \oplus (\textbf{1},\textbf{1})_{(-1;0;-6)} \dots & ,
\end{array}\notag
\ee
where the dots denote other irrelevant irreducible pieces which represent non-geometric fluxes in this frame.  
By means of the \eqref{dictionary_charges} and the relations \eqref{rhotausigma/STU}, the eight irrep's appearing above, can be instead recognised as the various internal components of $F_{(3)}$ and 
$H_{(3)}$ gauge fluxes. The corresponding flux-induced superpotential couplings are collected in table~\ref{Table:Fluxes_T6}.
\begin{table}[t!]
\begin{center}
\begin{tabular}{|c|c|c|c|}
\hline
STU couplings & Type IIB fluxes & Flux labels & $\mathbb{R}^{+}_{S} \times \mathbb{R}^{+}_{T} \times \mathbb{R}^{+}_{U} \times \textrm{SL}(3)_{a} \times \textrm{SL}(3)_{i}$ irrep's \\
\hline
\hline
$1$ & $F_{ijk}$ & $a_{0}$ & $(\textbf{1},\textbf{1})_{(+\frac{1}{2};+\frac{3}{2};+\frac{3}{2})}$ \\
\hline
$U$ & $F_{ajk}$ & $a_{1}$ & $(\textbf{3}^{\prime},\textbf{3})_{(+\frac{1}{2};+\frac{3}{2};+\frac{1}{2})}$\\
\hline
$U^{2}$ & $F_{abk}$ & $a_{2}$ & $(\textbf{3},\textbf{3}^{\prime})_{(+\frac{1}{2};+\frac{3}{2};-\frac{1}{2})}$ \\
\hline
$U^{3}$ & $F_{abc}$ & $a_{3}$ & $(\textbf{1},\textbf{1})_{(+\frac{1}{2};+\frac{3}{2};-\frac{3}{2})}$\\
\hline
$S$ & $H_{ijk}$ & $b_{0}$ & $(\textbf{1},\textbf{1})_{(-\frac{1}{2};+\frac{3}{2};+\frac{3}{2})}$\\
\hline
$S \, U$ & $H_{ajk}$ &  $b_{1}$ & $(\textbf{3}^{\prime},\textbf{3})_{(-\frac{1}{2};+\frac{3}{2};+\frac{1}{2})}$\\
\hline
$S \, U^{2}$  & $H_{abk}$  & $b_{2}$  & $(\textbf{3},\textbf{3}^{\prime})_{(-\frac{1}{2};+\frac{3}{2};-\frac{1}{2})}$\\
\hline
$S \, U^{3}$ & $H_{abc}$ & $ b_{3}$ & $(\textbf{1},\textbf{1})_{(-\frac{1}{2};+\frac{3}{2};-\frac{3}{2})}$ \\
\hline
\end{tabular}
\end{center}
\caption{{\it Summary of type IIB fluxes and superpotential couplings on a $\mathbb{T}^{6}$. Isotropy (\emph{i.e.} $\textrm{SO}(3)$-invariance) only allows for flux components that can be constructed by 
using $\epsilon_{(3)}$'s and $\delta_{(3)}$'s. 
These symmetries also induce a natural splitting $\,\eta^{m}=(\eta^{a} \,,\, \eta^{i} )\,$ where $\,a=1,3,5\,$ and $\,i=2,4,6\,$.}}
\label{Table:Fluxes_T6}
\end{table}
The explicit (isotropic) superpotential reads
\be
\label{W_T6}
\mathcal{W}_{(\mathbb{T}^{6})} \ = \ a_{0} \, - \, 3a_{1}U \, + \,3 a_{2}U^{2} \, - \, a_{3}U^{3} 
\, - \, S \,( b_{0} \, - \, 3b_{1} U \, + \,3 b_{2}U^{2} \, - \, b_{3}U^{3} ) \ .
\ee
One should note that the underlying gauging for this class of compactifications is \emph{Abelian}.
This is in line with what already observed in refs~\cite{Dibitetto:2011gm} when studying the connection between type IIB compactifications on a $\mathbb{T}^{6}$ with D$3$-branes and O$3$-planes as sources  
where the corresponding effective 4D description turned out to be $\mathcal{N}=4$ supergravity with $\textrm{U}(1)^{12}$ gauge group.

Finally, we want to check the (non-)BPS tadpoles that such backgrounds produce by plugging the corresponding $f_{\alpha MNP}$ into the QC \eqref{QC4} \& \eqref{QC8}. The $\mathcal{N}=4$ QC in \eqref{QC4}
 turn out to be trivially satisfied as they should, whereas \eqref{QC8} produces BPS-tadpoles of the form
\be
\begin{array}{lclc}
a_{3}b_{0}-3a_{2}b_{1}+3a_{1}b_{2}-a_{0}b_{3} & \equiv & N_{(\textrm{O}3/\textrm{D}3)} & ,
\end{array}
\ee
just as expected.

\subsection{Compactifications on $S^{3} \times S^{3}$}
\label{sec:S3S3}

In this case, one can still include 3-form fluxes both of NS-NS and R-R type, but restricted to those components which do not have mixed legs within $S^{3}_{a}$ \& $S^{3}_{i}$. This is due to the
special topological property of each $S^{3}$ of lacking non-trivial 1- and 2-cycles. Besides these gauge fluxes, the geometry of both 3-spheres is described by $3\,\times\,3$ symmetric matrices
$\Theta_{ab}$ and $\Theta_{ij}$ representing their metric in flat coordinates. 

We will use the same decomposition chain as in the toroidal case, but bearing in mind that $\Theta_{ab}$ and $\Theta_{ij}$ parametrising the internal curvature naturally come from $\textbf{10}'$'s of the
two intermediate $\textrm{SL}(4)$ factors, due to the natural embeding of $S^{3}$ into $\mathbb{R}^{4}$. This procedure yields
\be
\begin{array}{cclc}
(\textbf{2},\textbf{220})  & \rightarrow  &  (\textbf{1},\textbf{1})_{(+1;-6;0)}  \oplus (\textbf{1},\textbf{1})_{(-1;-6;0)} \oplus 
(\textbf{1},\textbf{1})_{(+1;0;-6)}  \oplus (\textbf{1},\textbf{1})_{(-1;0;-6)}  \oplus \\
 & & (\textbf{6}',\textbf{1})_{(+1;-2;0)}  \oplus (\textbf{6}',\textbf{1})_{(-1;-2;0)} \oplus 
(\textbf{1},\textbf{6}')_{(+1;0;-2)}  \oplus (\textbf{1},\textbf{6}')_{(-1;0;-2)} \oplus \dots & .
\end{array}\notag
\ee
The resulting set of superpotential couplings obtained in this way upon using \eqref{dictionary_charges}, is given in table~\ref{Table:Fluxes_S3S3}.
\begin{table}[t!]
\begin{center}
\begin{tabular}{|c|c|c|c|}
\hline
STU couplings & Type IIB fluxes & Flux labels & $\mathbb{R}^{+}_{S} \times \mathbb{R}^{+}_{T} \times \mathbb{R}^{+}_{U} \times \textrm{SL}(3)_{a} \times \textrm{SL}(3)_{i}$ irrep's \\
\hline
\hline
$1$ & $F_{ijk}$ & $a_{0}$ & $(\textbf{1},\textbf{1})_{(+\frac{1}{2};+\frac{3}{2};+\frac{3}{2})}$ \\
\hline
$U^{3}$ & $F_{abc}$ & $a_{3}$ & $(\textbf{1},\textbf{1})_{(+\frac{1}{2};+\frac{3}{2};-\frac{3}{2})}$\\
\hline
$S$ & $H_{ijk}$ & $b_{0}$ & $(\textbf{1},\textbf{1})_{(-\frac{1}{2};+\frac{3}{2};+\frac{3}{2})}$\\
\hline
$S \, U^{3}$ & $H_{abc}$ & $ b_{3}$ & $(\textbf{1},\textbf{1})_{(-\frac{1}{2};+\frac{3}{2};-\frac{3}{2})}$ \\
\hline
\hline
$T \, U $ & $\Theta^{(+)}_{ab}$ & $\tilde{c}_{1}$ & $(\textbf{6}^{\prime},\textbf{1})_{(+\frac{1}{2};+\frac{1}{2};+\frac{1}{2})}$ \\
\hline
$T \, U^{2} $ & $\Theta^{(+)}_{ij}$ & $\tilde{c}_{2}$ & $(\textbf{1},\textbf{6}^{\prime})_{(+\frac{1}{2};+\frac{1}{2};-\frac{1}{2})}$ \\
\hline
$S \, T \, U $ & $\Theta^{(-)}_{ab}$ & $\tilde{d}_{1}$ & $(\textbf{6}^{\prime},\textbf{1})_{(-\frac{1}{2};+\frac{1}{2};+\frac{1}{2})}$ \\
\hline
$S \, T \, U^{2} $ & $\Theta^{(-)}_{ij}$ & $\tilde{d}_{2}$ & $(\textbf{1},\textbf{6}^{\prime})_{(-\frac{1}{2};+\frac{1}{2};-\frac{1}{2})}$ \\
\hline
\end{tabular}
\end{center}
\caption{{\it Summary of type IIB fluxes and superpotential couplings on $S^{3}\times S^{3}$. 
Isotropy (\emph{i.e.} $\textrm{SO}(3)$-invariance) only allows for flux components that can be constructed by using $\epsilon_{(3)}$'s and $\delta_{(3)}$'s. 
Our chosen frame includes $F_{(3)}$ \& $H_{(3)}$ fluxes as $\Theta^{(\pm)}_{ab}$ \& $\Theta^{(\pm)}_{ij}$ describing the $S^{3}_{a}$ \& $S^{3}_{i}$ geometry, respectively.}}
\label{Table:Fluxes_S3S3}
\end{table}
The associated flux-induced superpotential is given by
\be
\label{W_S3S3}
\mathcal{W}_{(S^{3} \times S^{3})} \ = \ a_{0} \, - \, a_{3}U^{3} \, - \, S \, (b_{0} \, - \, b_{3}U^{3} ) \, - \, 3\tilde{c}_{1}TU \, + \, 3\tilde{c}_{2}TU^{2} 
\, + \, S \, (3\tilde{d}_{1}TU \, - \, 3\tilde{d}_{2}TU^{2}) \ .
\ee
The $\mathcal{N}=4$ QC \eqref{QC4} are trivially satisfied, thus always yielding a consistent half-maximal 4D supergravity with gauge group\footnote{We denote by $\textrm{ISO}(3)\,\equiv\,
\textrm{CSO}(3,0,1)$ the contracted version of $\textrm{SO}(4)$ describing the isometries of $\mathbb{R}^{3}$, consisting of $3$ rotations and $3$ translations.} 
$\textrm{ISO}(3)\,\times\,\textrm{ISO}(3)$ \cite{Dibitetto:2011gm}.
Instead, by plugging the embedding tensor into the \eqref{QC8}, one can realise that these backgrounds are generically supported by the following tadpole-induced sources
\be
\begin{array}{lclc}
a_{3}b_{0} \ - \ a_{0}b_{3} & \equiv & N_{(\textrm{O}3/\textrm{D}3)} & , \\[2mm]
\tilde{c}_{1}\tilde{d}_{2} \ - \ \tilde{c}_{2}\tilde{d}_{1} & \equiv & N_{(\textrm{??})} & , 
\end{array}
\ee
where the second of the above tadpoles should be viewed as a source of SUSY-breaking coming from geometry.  

A particularly \emph{simple} subcase of this is given by the ``KS-like'' situation \cite{Klebanov:2000hb} where, \emph{e.g.} $F_{(3)}$ is only wrapping $S^{3}_{i}$ and $H_{(3)}$ only $S^{3}_{a}$.
In such a situation, the corresponding superpotential reads
\be
\label{W_KS}
\mathcal{W}_{\textrm{(KS)}} \ = \ a_{0} \, - \,  b_{3}SU^{3} \, - \, 3\tilde{c}_{1}TU  \, - \, 3\tilde{d}_{2}STU^{2} \ .
\ee
Note that this STU-model can be reinterpreted as the one induced by a type IIB reduction on $\mathbb{T}^{6}$ with \emph{non-geometric} $Q$ \& $P$ fluxes given by
\be
\begin{array}{lcccclc}
{Q_{a}}^{bc} \ \equiv \ \Theta^{(+)}_{ad} \, \epsilon^{dbc} & , & & \textrm{and} & & {P_{i}}^{jk} \ \equiv \ \Theta^{(-)}_{il} \, \epsilon^{ljk} & ,
\end{array}
\ee
where $\epsilon^{abc}$ \& $\epsilon^{ijk}$ denote the $\textrm{SL}(3)_{a}$ \& $\textrm{SL}(3)_{i}$ Levi-Civita symbols, respectively.

It is of utmost interest to notice that these particular theories described by the superpotential \eqref{W_KS} were found in ref.~\cite{Dibitetto:2011gm} to possess (non-)supersymmetric AdS as well as unstable dS 
critical points. Here we propose the $S^{3} \times S^{3}$ compactification of type IIB as their 10D interpretation, which was previously lacking.

\subsection{Compactifications on $S^{3} \times \mathbb{T}^{3}$}
\label{sec:S3T3}

The allowed gauge fluxes in this case are exactly those ones that are also present in the $S^{3} \times S^{3}$ model. For what concerns the curvature, everything within $S^{3}_{a}$ remains unchanged
w.r.t. what can be found in table~\ref{Table:Fluxes_S3S3}, while no curvature flux is present in $\mathbb{T}^{3}_{i}$. 
This means that we do not need to perform any new group-theoretical branchings in order to derive the underlying flux-induced superpotential for such a model. It can simply be obtained from
\eqref{W_S3S3} by setting $\tilde{c}_{2}\,= \,\tilde{d}_{2}\,= \,0$; this yields
\be
\label{W_S3T3}
\mathcal{W}_{(S^{3} \times \mathbb{T}^{3})} \ = \ a_{0} \, - \, a_{3}U^{3} \, - \, S \, (b_{0} \, - \, b_{3}U^{3} ) \, - \, 3\tilde{c}_{1}TU \, + \, 3\tilde{d}_{1}STU  \ .
\ee

Also in this case, the $\mathcal{N}=4$ QC are trivially satisfied, implying that \eqref{W_S3T3} always describes an orientifold reduction of type IIB preserving sixteen supercharges.
Moreover, the $\mathcal{N}=8$ QC show that the only BPS extended objects supporting such backgrounds are D$3$-branes and O$3$-planes: 
\be
\begin{array}{lclc}
a_{3}b_{0}-3a_{2}b_{1}+3a_{1}b_{2}-a_{0}b_{3} & \equiv & N_{(\textrm{O}3/\textrm{D}3)} & ,
\end{array}
\ee
whereas $N_{(\textrm{??})} \, = \, 0$. It may be worth mentioning that this class of effective theories, which is interesting in itself, has not been studied in detail and in particular it still remains
to be seen whether it admits maximally symmetric vacua or it just describes warped backgrounds in type IIB.

\section{Discussion}
\label{sec:Discussion}

In this paper we have studied some features of (non-)toroidal backgrounds of type IIB superstring theory allowing for four-dimensional gauged supergravity models as effective descriptions.
In particular, we focused on examples with spacetime-filling orientifold planes thus preserving sixteen supercharges in connection with $\mathcal{N}=4$ supergravities in $D=4$.
A suitable truncation to the isotropic sector of these theories turns out to be described by minimal STU-models with superpotential deformations to be interpreted as generalised fluxes.

In this context, by choosing $\mathbb{T}^{6}$ as a standard reference background, most of the superpotential couplings will turn out to correspond to \emph{non-geometric} flux deformations thereof. 
Nevertheless, inspired by the philosophy of ref.~\cite{Danielsson:2015tsa}, we make use of group-theoretical arguments within a ``bottom-up'' approach in order to conclude that particular sets of
would-be non-geometric fluxes in fact just correspond to having considered different backgrounds other than toroidal as a starting point.

The main result of our present analysis is the prediction of the possibility of breaking the no-scale symmetry, typical of type IIB toroidal reductions with gauge fluxes and preventing one to perturbatively 
lift the K\"ahler moduli, by just considering fluctuations around $S^{3}\times S^{3}$ or $S^{3}\times \mathbb{T}^{3}$ rather than $\mathbb{T}^{6}$. The explicit flux-induced superpotentials are given,
and the BPS extended objects supporting these backgrounds are discussed. The explicit 10D construction proving the existence of a consistent truncation of type IIB supergravity on $S^{3}\times S^{3}$
still remains to be worked out, but we leave it for future work \cite{AdS4S3S3}.

In conclusion, our results indicate a novel path to be pursued in the context of type IIB flux compactifications, and, possibly, de Sitter model-building. More specifically, note that the superpotential
\eqref{W_S3S3} describes a model where O-planes are present together with the possibility of having negative sectional curvature, thus circumventing the no-go theorem in ref.~\cite{Maldacena:2000mw}.
The final scope of such a programme could be that of having access to constructions as those ones presented in \cite{Blaback:2015zra} generically yielding stable de Sitter solutions in $\mathcal{N}=1$ 
STU-models, but now with superpotentials that can be made geometric in the sense explained here.

%
%

\appendix

\section{Relevant branching rules}
\label{sec:app1}

In this appendix we collect the whole set of branching rules used in the present paper. We refer to \cite{Feger:2012bs} for the conventions adopted
here.

\be
\begin{array}{cclc}
\textrm{E}_{7(7)}  & \supset  & \textrm{SL}(2) \times \textrm{SO}(6,6) & , \\[3mm]
\textbf{56}  & \rightarrow  & (\textbf{2},\textbf{12})  \oplus (\textbf{1},\textbf{32})  & ,\\[2mm]
\textbf{133}  & \rightarrow  & (\textbf{3},\textbf{1})  \oplus (\textbf{1},\textbf{66}) \oplus (\textbf{2},\textbf{32}^{\prime}) & ,\\[2mm]
\textbf{912}  & \rightarrow  & (\textbf{2},\textbf{12})  \oplus (\textbf{2},\textbf{220}) \oplus (\textbf{3},\textbf{32}) \oplus (\textbf{1},\textbf{352}^{\prime}) & .
\end{array}\label{E7->SL2SO66}
\ee

\be
\begin{array}{cclc}
\textrm{SO}(6,6)  & \supset  & \textrm{SL}(4) \times \textrm{SL}(4)  & , \\[3mm]
\textbf{12}  & \rightarrow  & (\textbf{6},\textbf{1})  \oplus (\textbf{1},\textbf{6}) & ,\\[2mm]
\textbf{32}  & \rightarrow  & (\textbf{4},\textbf{4}^{\prime}) \oplus (\textbf{4}^{\prime},\textbf{4}) & ,\\[2mm]
\textbf{66}  & \rightarrow  & (\textbf{15},\textbf{1})  \oplus (\textbf{1},\textbf{15}) \oplus (\textbf{6},\textbf{6}) & ,\\[2mm]
\textbf{220}  & \rightarrow  & (\textbf{10},\textbf{1})  \oplus (\textbf{10}^{\prime},\textbf{1}) \oplus (\textbf{1},\textbf{10})  \oplus (\textbf{1},\textbf{10}^{\prime}) 
\oplus (\textbf{6},\textbf{15})  \oplus (\textbf{15},\textbf{6}) & .
\end{array}
\ee

\be
\begin{array}{cclc}
\textrm{SL}(4)  & \supset  & \mathbb{R}^{+} \times \textrm{SL}(3)  & , \\[3mm]
\textbf{4}  & \rightarrow  & \textbf{1}_{(-3)}  \oplus \textbf{3}_{(+1)}  & ,\\[2mm]
\textbf{6}  & \rightarrow  & \textbf{3}_{(-2)}  \oplus \textbf{3}^{\prime}_{(+2)} & ,\\[2mm]
\textbf{10}  & \rightarrow  & \textbf{1}_{(-6)}  \oplus \textbf{3}_{(-2)} \oplus \textbf{6}_{(+2)} & ,\\[2mm]
\textbf{15}  & \rightarrow  & \textbf{1}_{(0)}  \oplus \textbf{3}_{(+4)} \oplus \textbf{3}^{\prime}_{(-4)} \oplus \textbf{8}_{(0)} & ,\\[2mm]
\textbf{20}  & \rightarrow  & \textbf{3}_{(+1)}  \oplus \textbf{3}^{\prime}_{(+5)}  \oplus \textbf{6}^{\prime}_{(+1)} \oplus \textbf{8}_{(-3)} & ,
\end{array}
\ee
where the subscripts in the above decompisotions denote $\mathbb{R}^{+}$ charges.

\be
\begin{array}{cclc}
\textrm{SL}(2)  & \supset  & \mathbb{R}^{+}  & , \\[3mm]
\textbf{2}  & \rightarrow  & \textbf{1}_{(-1)}  \oplus \textbf{1}_{(+1)}  & ,\\[2mm]
\textbf{3}  & \rightarrow  & \textbf{1}_{(-2)}  \oplus \textbf{1}_{(0)} \oplus \textbf{1}_{(+2)} & ,\\[2mm]
\textbf{4}  & \rightarrow  & \textbf{1}_{(-3)}  \oplus \textbf{1}_{(-1)} \oplus \textbf{1}_{(+1)} \oplus \textbf{1}_{(+3)} & ,
\end{array}
\ee
where the subscripts in the above decompisotions denote $\mathbb{R}^{+}$ charges.

\section{Type IIB fluxes and the embedding tensor $\,f_{\alpha MNP}$}
\label{sec:app2}

In this appendix, we summarise the identification between embedding tensor components $\,f_{\a MNP}\,$ in the $(\textbf{2},\textbf{220})$ (alternatively $\, \Lambda_{\a ABC}\,$ as explained in 
section~\ref{sec:truncation}) and type IIB flux backgrounds for the $\,\mathcal{N}=1\,$ supergravity theory.

\begin{figure}[h!]
\begin{center}
\scalebox{0.9}[0.9]{
\begin{tabular}{ccccc}
\includegraphics[scale=0.5,keepaspectratio=true]{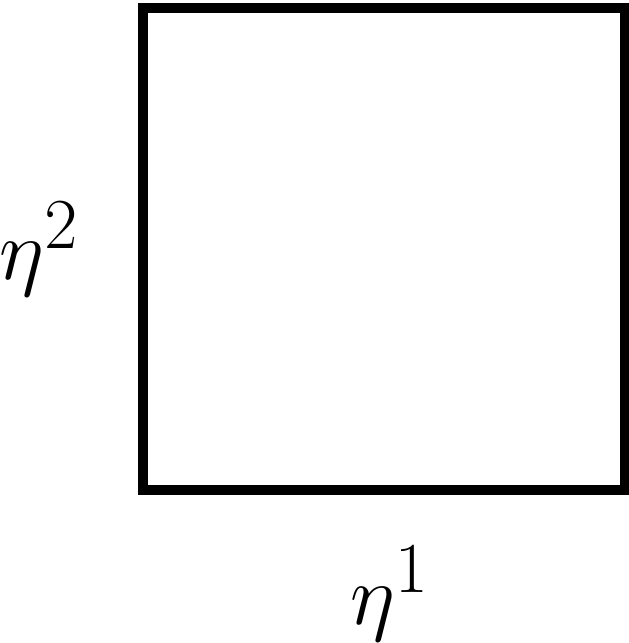} &    &  \includegraphics[scale=0.5,keepaspectratio=true]{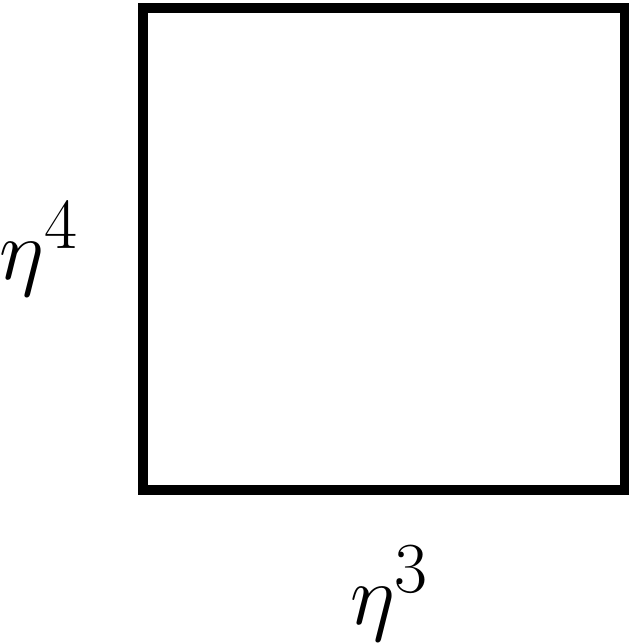}  &    & \includegraphics[scale=0.5,keepaspectratio=true]{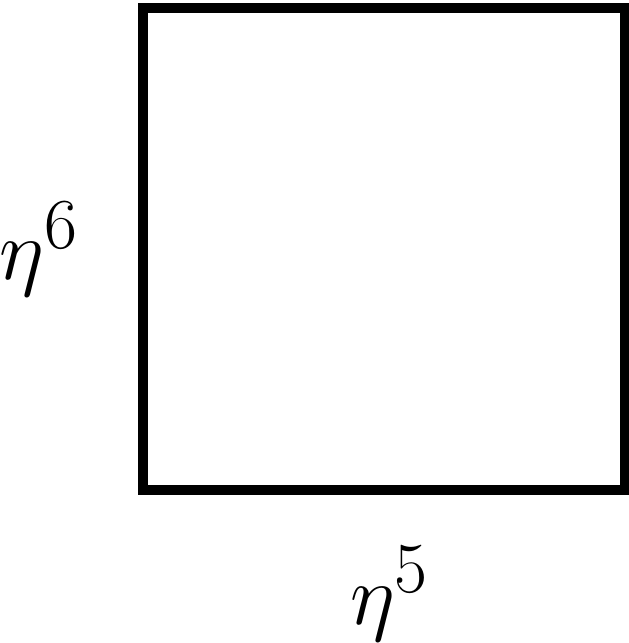} \\[-24mm]
     & \,\,\Large{$\times$} &        & \,\,\Large{$\times$} &
\end{tabular}
}
\end{center}
\vspace{9.5mm}
\caption{{\it $\mathbb{T}^{6} =  \mathbb{T}^{2}_{1} \times \mathbb{T}_{2}^{2} \times \mathbb{T}_{3}^{2}$ torus factorisation and the coordinate basis.}}
\label{fig:Torus_Factor1}
\end{figure}
In the following we will use early Latin indices $a,b,c$ for horizontal $\,``-"$ $x$-like directions $(\eta^{1},\eta^{3},\eta^{5})$ and late Latin indices $i,j,k$ for vertical $\,``\,|\,"$ $y$-like 
directions $(\eta^{2},\eta^{4},\eta^{6})$ in the 2-tori $\,T_{I}\,$ with $\,I=1,2,3$. This splitting of coordinates is in one-to-one correspondence with the SO($6,6$) index splitting of the embedding 
tensor components given in (\ref{ET_splitting}), where $A=(1,2,3,4) \equiv (a,i,\bar{a},\bar{i})\,$ refers to an SO($2,2$) fundamental index and $\epsilon_{IJK}$ denotes the usual totally antisymmetric 
Levi-Civita tensor. 
\begin{table}[h!]
\renewcommand{\arraystretch}{1.25}
\begin{center}
\scalebox{0.84}[0.87]{
\begin{tabular}{ | c || c | c | c | c |}
\hline
couplings & SO($6,6$) & SO($2,2$) & type IIB & fluxes\\
\hline
\hline
$1 $& $ -f_{+ \bar{a}\bar{b}\bar{c}} $  & $ - \Lambda_{+333} $& $ {F}_{ ijk} $&  $  a_0 $\\
\hline
$U $& $f_{+ \bar{a}\bar{b}\bar{k}}$  &  $ \Lambda_{+334} $& ${F}_{ ij c} $&  $   a_1 $\\
\hline
$U^2 $& $ -f_{+ \bar{a}\bar{j}\bar{k}}$  & $ - \Lambda_{+344} $& ${F}_{i b c} $&  $  a_2 $\\
\hline
$U^3 $& $f_{+ \bar{i}\bar{j}\bar{k}}$  & $ \Lambda_{+444} $& ${F}_{a b c} $&  $  a_3 $\\
\hline
\hline
$S $& $ -f_{- \bar{a}\bar{b}\bar{c}} $  &  $ - \Lambda_{-333} $& $ {H}_{ijk} $&  $  - b_0$\\
\hline
$S \, U $& $f_{- \bar{a}\bar{b}\bar{k}}$  & $ \Lambda_{-334}$ & $ {H}_{ij c} $&  $  - b_1 $\\
\hline
$S \, U^2 $& $ -f_{- \bar{a}\bar{j}\bar{k}}$   & $ - \Lambda_{-344} $& $ {H}_{ i b c}$ &  $ - b_2 $\\
\hline
$S \, U^3 $& $f_{- \bar{i}\bar{j}\bar{k}}$   &  $ \Lambda_{-444} $& $ {H}_{a b c} $&  $ - b_3 $\\
\hline
\hline
$T $& $f_{+ \bar{a}\bar{b}k}$   & $ \Lambda_{+233} $& $  Q^{a b}_k $ & $  c_0 $\\
\hline
$T \, U $& $f_{+ \bar{a}\bar{j} k}=f_{+ \bar{i}\bar{b} k}\,\,\,,\,\,\,f_{+ a\bar{b}\bar{c}}$  &  $ \Lambda_{+234} \,\,\,,\,\,\, \Lambda_{+133} $& $ Q ^{a j}_k = Q^{i b}_k \,\,\,,\,\,\, Q^{b c}_a $&  $c_1 \,\,\,,\,\,\, \tilde {c}_1 $\\
\hline
$T \, U^2 $& $f_{+ \bar{i}\bar{b}c}=f_{+ \bar{a}\bar{j}c}\,\,\,,\,\,\,f_{+ \bar{i}\bar{j}k}$   &  $ \Lambda_{+134} \,\,\,,\,\,\, \Lambda_{+244} $& $ Q ^{ib}_c = Q^{a j}_c \,\,\,,\,\,\, Q^{ij}_k $&  $c_2 \,\,\,,\,\,\,\tilde{c}_2 $\\
\hline
$T \, U^3 $& $f_{+ \bar{i}\bar{j} c}$  &  $ \Lambda_{+144} $& $  Q^{ij}_{c} $&  $c_3 $\\
\hline
\hline
$S \, T $& $f_{- \bar{a}\bar{b}k}$   &  $ \Lambda_{-233} $& $  P^{a b}_k $&  $  - d_0 $\\
\hline
$S \, T \, U $& $f_{- \bar{a}\bar{j} k}=f_{- \bar{i}\bar{b} k}\,\,\,,\,\,\,f_{- a\bar{b}\bar{c}}$   &  $ \Lambda_{-234} \,\,\,,\,\,\, \Lambda_{-133} $& $ P ^{a j}_k = P^{i b}_k \,\,\,,\,\,\, P^{b c}_a $&   $- d_1 \,\,\,,\,\,\, - \tilde {d}_1 $\\
\hline
$S \, T \, U^2 $& $f_{- \bar{i}\bar{b}c}=f_{- \bar{a}\bar{j}c}\,\,\,,\,\,\,f_{- \bar{i}\bar{j}k}$   &  $ \Lambda_{-134} \,\,\,,\,\,\, \Lambda_{-244} $& $ P ^{ib}_c = P^{a j}_c \,\,\,,\,\,\, P^{ij}_k $&   $ - d_2 \,\,\,,\,\,\, - \tilde{d}_2 $\\
\hline
$S \, T \, U^3 $& $f_{- \bar{i}\bar{j} c}$  &  $ \Lambda_{-144} $ & $  P^{ij}_{c} $&   $ - d_3 $\\
\hline
\end{tabular}
}
\end{center}
\caption{{\it Mapping between unprimed fluxes, embedding tensor components and couplings in the flux-induced superpotential. We have made the index splitting $\,M=\{a,i,\bar{a},\bar{i}\}\,$ for 
$\textrm{SO}(6,6)$ light-cone coordinates.}}
\label{table:unprimed_fluxes4}
\end{table}

The dictionary between embedding tensor components and type IIB generalised fluxes can be found here in tables~\ref{table:unprimed_fluxes4} and \ref{table:primed_fluxes4}.
Such an identification was originally proposed in ref.~\cite{Dibitetto:2010rg} and further developed in ref.~\cite{Dibitetto:2011gm}. 
\begin{table}[h!]
\renewcommand{\arraystretch}{1.25}
\begin{center}
\scalebox{0.87}[0.87]{
\begin{tabular}{ | c || c | c | c | c |}
\hline
couplings & SO($6,6$) & SO($2,2$) & type IIB &  fluxes\\
\hline
\hline
$T^3 \, U^3 $& $ -f_{+ abc} $ & $ - \Lambda_{+111} $& $ {F'}^{ijk} $&   $  a_0' $\\
\hline
$T^3 \, U^2 $&  $f_{+ abk}$ &  $ \Lambda_{+112} $& ${F'}^{ ij c} $& $   a_1' $\\
\hline
$T^3 \, U $& $ -f_{+ ajk}$  & $ - \Lambda_{+122} $& ${F'}^{i b c} $& $  a_2' $\\
\hline
$ T^3 $& $f_{+ ijk}$ & $ \Lambda_{+222} $& ${F'}^{a b c} $& $  a_3' $\\
\hline
\hline
$S \, T^3 \, U^3 $& $ -f_{- abc} $  & $ - \Lambda_{-111} $& $ {H'}^{ ijk} $& $  - b_0'$\\
\hline
$S \, T^3 \, U^2 $& $f_{- abk}$  & $ \Lambda_{-112} $& $ {H'}^{i jc} $& $ - b_1' $\\
\hline
$S \, T^3 \, U $& $ -f_{- ajk}$  & $ - \Lambda_{-122} $& $ {H'}^{ i b c} $&  $ - b_2' $\\
\hline
$S  \, T^3 $& $f_{- ijk}$   & $ \Lambda_{-222}$ & $ {H'}^{a b c} $& $ - b_3' $\\
\hline
\hline
$T^2 \, U^3 $& $f_{+ ab\bar{k}}$ & $ \Lambda_{+114} $& $  {Q'}_{a b}^k $& $  c_0' $\\
\hline
$T^2 \, U^2 $& $f_{+ aj\bar{k}}=f_{+ ib\bar{k}}\,\,\,,\,\,\,f_{+ \bar{a}bc}$ & $  \Lambda_{+124} \,\,\,,\,\,\, \Lambda_{+113} $ & $ {Q'}_{a j}^k = {Q'}_{i b}^k \,\,\,,\,\,\, {Q'}_{b c}^a $& $c_1' \,\,\,,\,\,\, \tilde{c}_1' $\\
\hline
$T^2 \, U $& $f_{+ ib\bar{c}}=f_{+ aj\bar{c}}\,\,\,,\,\,\,f_{+ ij\bar{k}}$  & $ \Lambda_{+123} \,\,\,,\,\,\, \Lambda_{+224} $& $ {Q'}_{ib}^c = {Q'}_{a j}^c \,\,\,,\,\,\, {Q'}_{ij}^k $& $c_2' \,\,\,,\,\,\,\tilde{c}_2' $\\
\hline
$T^2 $& $f_{+ ij\bar{c}}$  & $ \Lambda_{+223} $& $  {Q'}_{ij}^{c} $& $c_3' $\\
\hline
\hline
$S \, T^2 \, U^3$& $f_{- ab\bar{k}}$  & $ \Lambda_{-114} $& $  {P'}_{a b}^k $& $ - d_0' $\\
\hline
$S \, T^2 \, U^2 $& $f_{- aj\bar{k}}=f_{- ib\bar{k}}\,\,\,,\,\,\,f_{- \bar{a}bc}$ & $ \Lambda_{-124} \,\,\,,\,\,\, \Lambda_{-113} $& $ {P'}_{a j}^k = {P'}_{i b}^k \,\,\,,\,\,\, {P'}_{b c}^a $& $ - d_1' \,\,\,,\,\,\, - \tilde {d}_1' $\\
\hline
$S \, T^2 \, U $& $f_{- ib\bar{c}}=f_{- aj\bar{c}}\,\,\,,\,\,\,f_{- ij\bar{k}}$  &$ \Lambda_{-123} \,\,\,,\,\,\, \Lambda_{-224} $ & $ {P'}_{ib}^c = {P'}_{a j}^c \,\,\,,\,\,\, {P'}_{ij}^k $& $ - d_2' \,\,\,,\,\,\, - \tilde{d}_2' $\\
\hline
$S \, T^2  $& $f_{-ij\bar{c}}$ & $ \Lambda_{-223} $ & $  {P'}_{ij}^{c} $& $ - d_3' $\\
\hline
\end{tabular}
}
\end{center}
\caption{{\it Mapping between primed fluxes, embedding tensor components and couplings in the flux-induced superpotential. We have made the index splitting $\,M=\{a,i,\bar{a},\bar{i}\}\,$ for 
$\textrm{SO}(6,6)$ light-cone coordinates.}}
\label{table:primed_fluxes4}
\end{table}

Irrespective of their string theory interpretation, the above set of fluxes generates the following $\,\mathcal{N}=1\,$ flux-induced superpotential given in \eqref{W_fluxes4}, involving the 
three complex moduli $S$, $T$ and $U$ surviving the SO($3$) truncation introduced in section~\ref{sec:truncation}. 
In the type IIB picture, the superpotential in (\ref{W_fluxes4}) contains flux-induced polynomials depending on both electric and magnetic pairs -- 
schematically $\,(e,m)\,$ -- of gauge $(F_{(3)},H_{(3)})$ fluxes and non-geometric $(Q,P)$ fluxes, as well as those induced by their less known primed counterparts $\,(F'_{(3)},H'_{(3)})\,$ and 
$\,(Q',P')\,$ fluxes.

%
%

\small

\clearpage

\bibliography{references}
\bibliographystyle{utphys}

\end{document}